\documentstyle[epsf]{aipproc}

\def\cwcplotmacro#1#2#3#4#5#6#7{\centering \leavevmode
    \vbox to#2{\rule{0pt}{#2}}
    \includegraphics{#1}}

\begin{document}

\title{QSO Absorption Lines: The UV Rest--Frame from $0<z<4$}

\author{Christopher W. Churchill}

\address{Department of Astronomy and Astrophysics \\ 
         The Pennsylvania State University \\
         University Park, PA 16802}

\maketitle

\begin{abstract}
By charting the kinematic, chemical, and ionization conditions of
galactic and intergalactic gas over the redshift range $0 - 4$ with
QSO absorption lines, the evolution of chemical abundances, the UV
meta--galactic background, and the clustering dynamics of galactic gas
can be studied.  
HIRES/Keck {{\rm Mg}\kern 0.1em{\sc ii}} $\lambda 2796$ profiles
arising in $z \sim 1$ galaxies are presented and the {{\rm Mg}\kern
0.1em{\sc ii}} kinematic clustering function is given.
The intriguing  $z=0.93$ systems toward Q1206+459 are shown and compared
to $z \sim 2$ HIRES/Keck {{\rm C}\kern 0.1em{\sc iv}} profiles to
illustrate how STIS/HST can be exploited for studies of the high
ionization conditions in $z \leq 1$ {{\rm Mg}\kern 0.1em{\sc ii}}
selected systems.   
The scientific motives and plans for a large IR $2 \leq z \leq 4$
{{\rm Mg}\kern 0.1em{\sc ii}} survey with the Hobby--Eberly Telescope
are presented.
\end{abstract}

\section*{Introduction}

Though QSO absorption lines have contributed many impressive strides
in our understanding of the UV universe (Charlton, this volume),
{\it terra--incognita\/} remains to be explored and scientifically
exploited.
A few of the ultimate aims of QSO absorption line studies are to
establish the history of cosmic chemical evolution, the shape and
intensity of the UV background, the evolving rate of galactic
accretion events and star bursting outflows, and the reciprocative
roles these play in galactic formation and evolution over $\sim 95$\%
of the age of the universe.
The UV rest--frame gaseous conditions seen in absorption at high
redshift do {\it not\/} suffer cosmologically induced effects that
might otherwise be mistaken as evolutionary processes
(i.e.~k--corrections).
Thus, from $z \sim 4$ to $z = 0$, statistical changes in absorbing
conditions, such as profile velocity spreads, numbers of
subcomponents, and ionization levels, can be unambiguously attributed
to evolution in either the structures, dynamics, numbers, and/or
ionization and chemical conditions of UV flux sensitive gas--phase
baryons.
In this contribution, the unknown ``absorbing'' UV universe is
discussed and scientific motives for its exploration are given. 

Neutral gas is easily traced using the {{\rm Ly}\kern 0.1em$\alpha$}
$\lambda 1216$ transition.
Traditionally, the {{\rm Mg}\kern 0.1em{\sc ii}} $\lambda\lambda 2796,
2803$ and the {{\rm C}\kern 0.1em{\sc iv}} $\lambda\lambda 1548,
1550$ resonant doublets have been used as tracers of the low and high
ionization gas, respectively, because they are very strong in
absorption and have easily identified doublet patterns.

\begin{figure}[hbt]
\cwcplotmacro{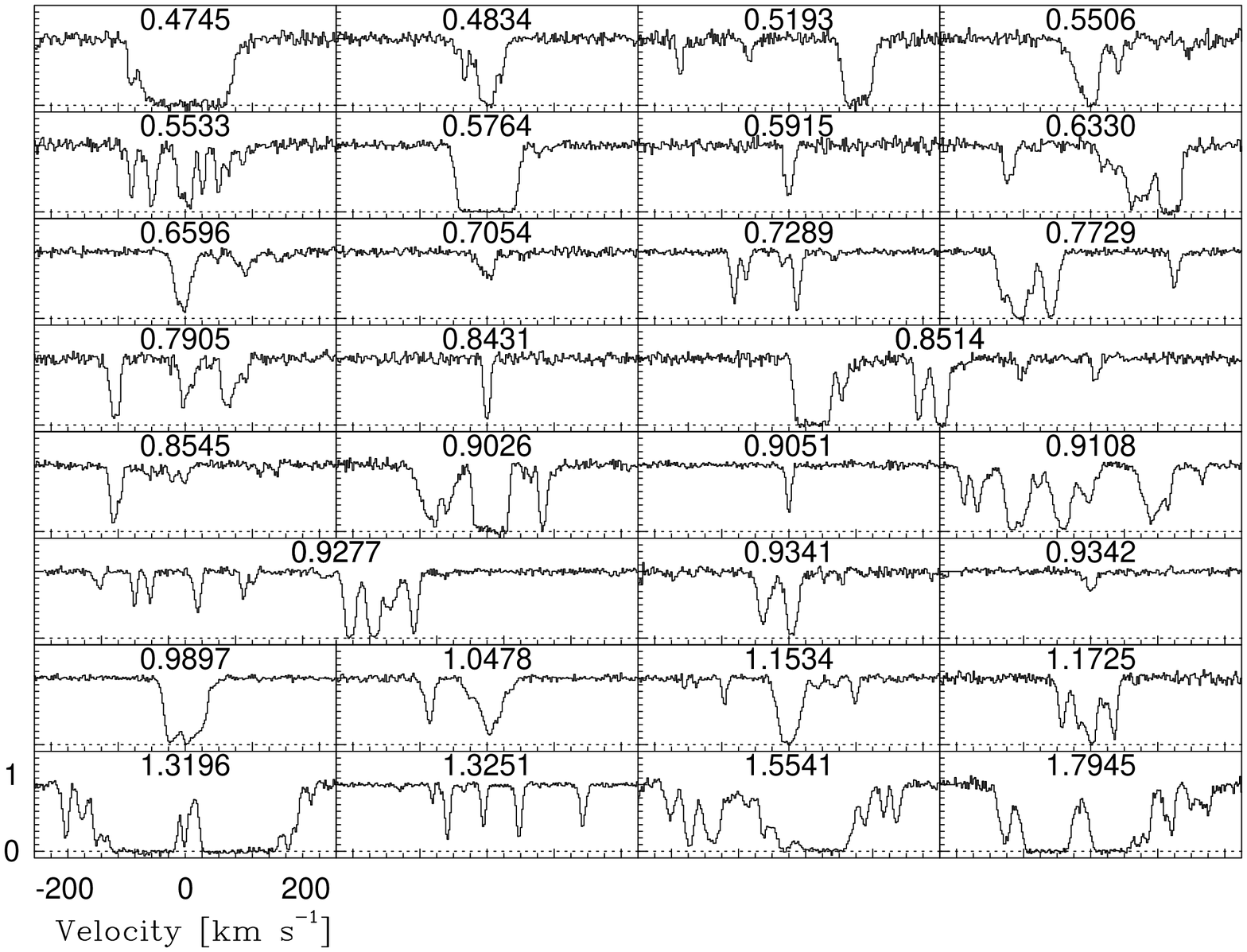}{3.5in}{0}{63.}{55.}{-256}{-28}
\caption{The $z < 2$ HIRES/Keck {{\rm Mg}\kern 0.1em{\sc ii}}
$\lambda$2796 transition in absorption presented in line of sight
velocity of the rest frame and in order of increasing redshift (marked
above the continuum).  The absorption arises in the extended low
ionization gas surrounding galaxies.  For each galaxy, the
$\lambda$2803 transition and several {{\rm Fe}\kern 0.1em{\sc ii}}
transitions have also been observed. 
\protect\label{cwcfig:hiresdata}}
%\vglue -0.24in
\end{figure}

Presented in Figure~\ref{cwcfig:hiresdata} are the {{\rm Mg}\kern
0.1em{\sc ii}} $\lambda 2796$ profiles from a HIRES/Keck survey
\cite{cwcref:thesis}.  
For 15 of these systems, Churchill {et.~al} \cite{cwcref:csv97}
compared the absorption and luminous properties of the galaxies.
They concluded that {{\rm Mg}\kern 0.1em{\sc ii}} absorbing gas
exhibits no clear spatial distribution or systematic kinematics and
suggested the gas results from episodic galactic processes.
The implications for galaxy evolution are not entirely clear.
The unexplored kinematics of the high ionization gas 
(i.e.~{{\rm C}\kern 0.1em{\sc iv}}) in these systems will be key for
further interpretation.

\begin{figure}[hbt]
\cwcplotmacro{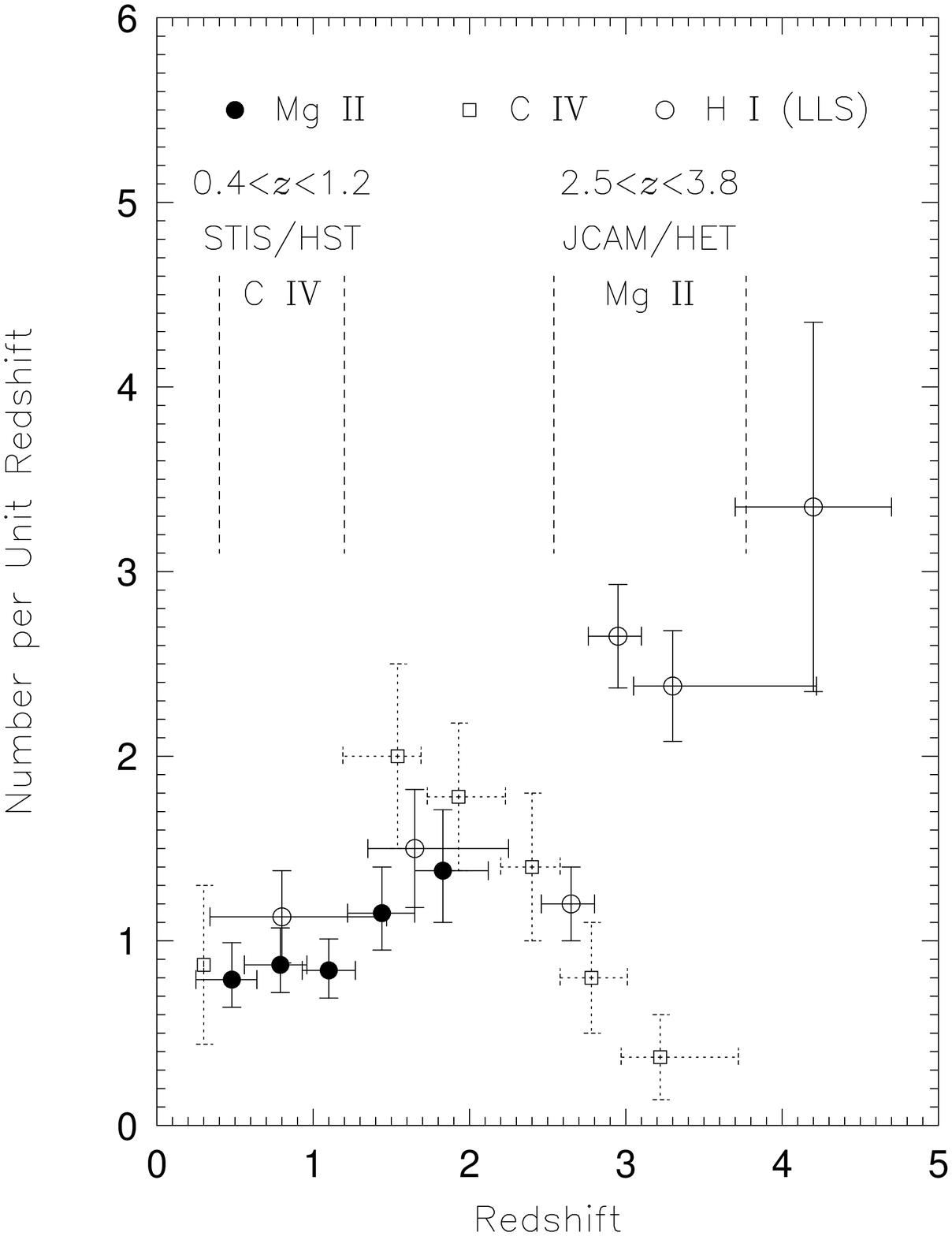}{1.85in}{0}{35.}{35.}{0}{-106}
\cwcplotmacro{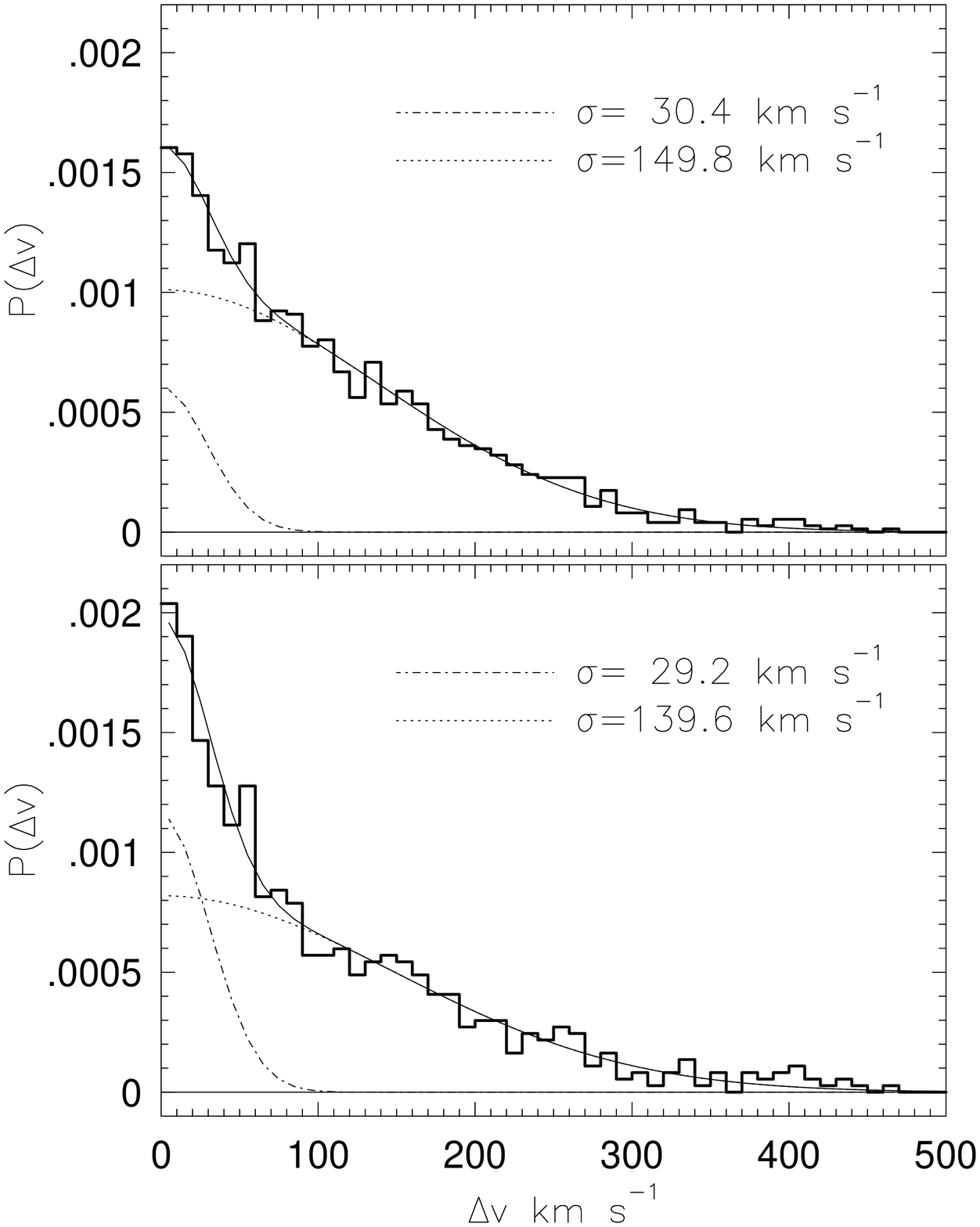}{1.85in}{0}{35.}{35.}{-203}{43}
\vglue -0.72in
\caption{
(left panels) The Two Point Clustering Function of {{\rm Mg}\kern
0.1em{\sc ii}} clouds in galactic halos.  The upper panel is for $0.4
\leq z \leq 1.7$, and the lower panel is for  $0.4 \leq z \leq 1.0$,
the latter presented because the higher redshift subsample is biased
toward the strongest absorption strengths. 
--- (right panel) The number per unit redshift for tracers of neutral
({{\rm H}\kern 0.1em{\sc i}}), low ({{\rm Mg}\kern 0.1em{\sc ii}}),
and high ({{\rm C}\kern 0.1em{\sc iv}}) ionization gas (adapted from
Steidel 1993 \protect\cite{cwcref:steidel93}).  
The unexplored UV absorbing universe and the required technology are
shown.
\protect\label{cwcfig:tpcf_dndz}}
\end{figure}

Shown in the left panels of Figure~\ref{cwcfig:tpcf_dndz} are the {{\rm
Mg}\kern 0.1em{\sc ii}} Two--Point Clustering Functions (TPCFs) of
subcomponents (blended ``clouds'' decomposed with Voigt Profiles).
The TPCF gives the probability of finding two clouds separated by
$\Delta v$ in an absorbing system \cite{cwcref:cvc97}.
In principle, parameterizing the TPCF by multi--component Gaussians
provides a means for statistically quantifying kinematic
evolution by comparing different redshift regimes.
Additionally, the TPCFs of low and high ionization gas can be compared
at similar epochs for quantifying the both relative ionization and
kinematic conditions.

The high resolution spectra required to study either {{\rm Mg}\kern
0.1em{\sc ii}} or {{\rm C}\kern 0.1em{\sc iv}} kinematic evolution
from $0 \leq z \leq 4$ do not exist-- nor do the spectra to study
their relative kinematics at $z\leq 1.2$ and $z\geq 2.2$.  
The former represents 50--70\% of the age of the universe and the
latter covers the epoch when galaxy evolution is believed to be 
very active.
What data exist, and what remains unknown?

The numbers of neutral, low, and high ionization systems are shown in
the right panel of Figure~\ref{cwcfig:tpcf_dndz}.
These numbers are the product of system number density and
gas cross section (ionization and metallicity dependent).
{{\rm Mg}\kern 0.1em{\sc ii}} has been surveyed from $0.3 < z < 2.2$
and traces the number of Lyman limit systems (LLS).
{{\rm C}\kern 0.1em{\sc iv}} has been surveyed from $1.2 < z < 3.7$
and shows a dramatic increase in the number of strong systems.
The HST Key Project will soon provide overlap between {{\rm C}\kern
0.1em{\sc iv}} and {{\rm Mg}\kern 0.1em{\sc ii}} for $z \leq 1.2$ (low
resolution only).
The $z \leq 1.2$ {{\rm C}\kern 0.1em{\sc iv}} kinematics are unknown,
and {{\rm Mg}\kern 0.1em{\sc ii}} is unexplored for $z \geq 2.2$.
High resolution spaced--based (UV) spectra are needed to extend {{\rm
C}\kern 0.1em{\sc iv}} kinematic studies to $z \leq 1.2$.
Near Infrared (IR) spectra are needed for a {\it complete
and uniform\/} survey of {{\rm Mg}\kern 0.1em{\sc ii}} to $z \geq
2.2$, followed by a higher resolution study of the gas kinematics.
These data are necessary if we are to establish a complete picture of
absorbing gas to $z=4$ for tracking the UV background, chemical
evolution, and inferring their roles in galactic evolution.

\section*{Detailing High Ionization at $z \leq 1.2$}

\begin{figure}[hbt]
\cwcplotmacro{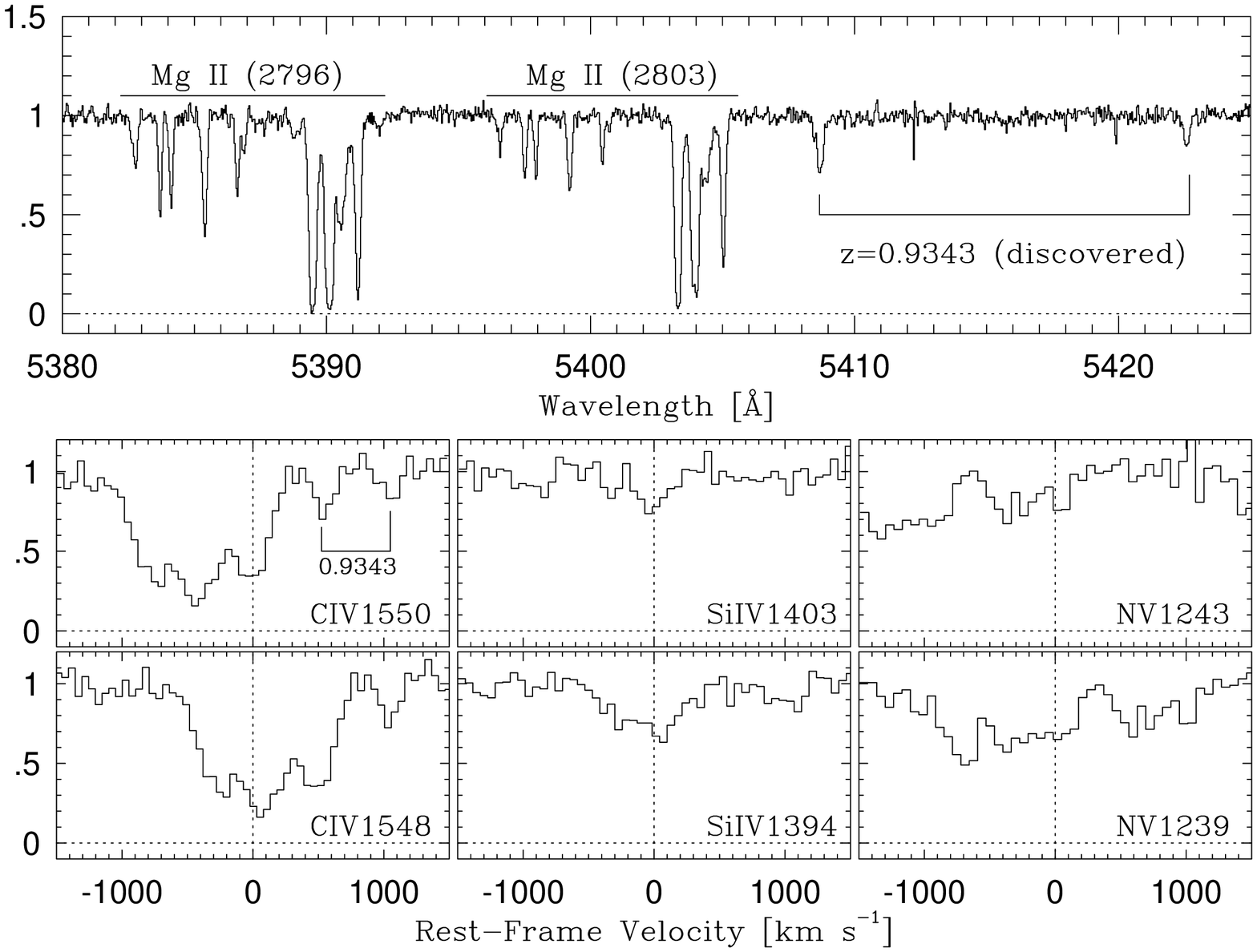}{1.7in}{0}{42.}{42.}{-108}{-90}
\cwcplotmacro{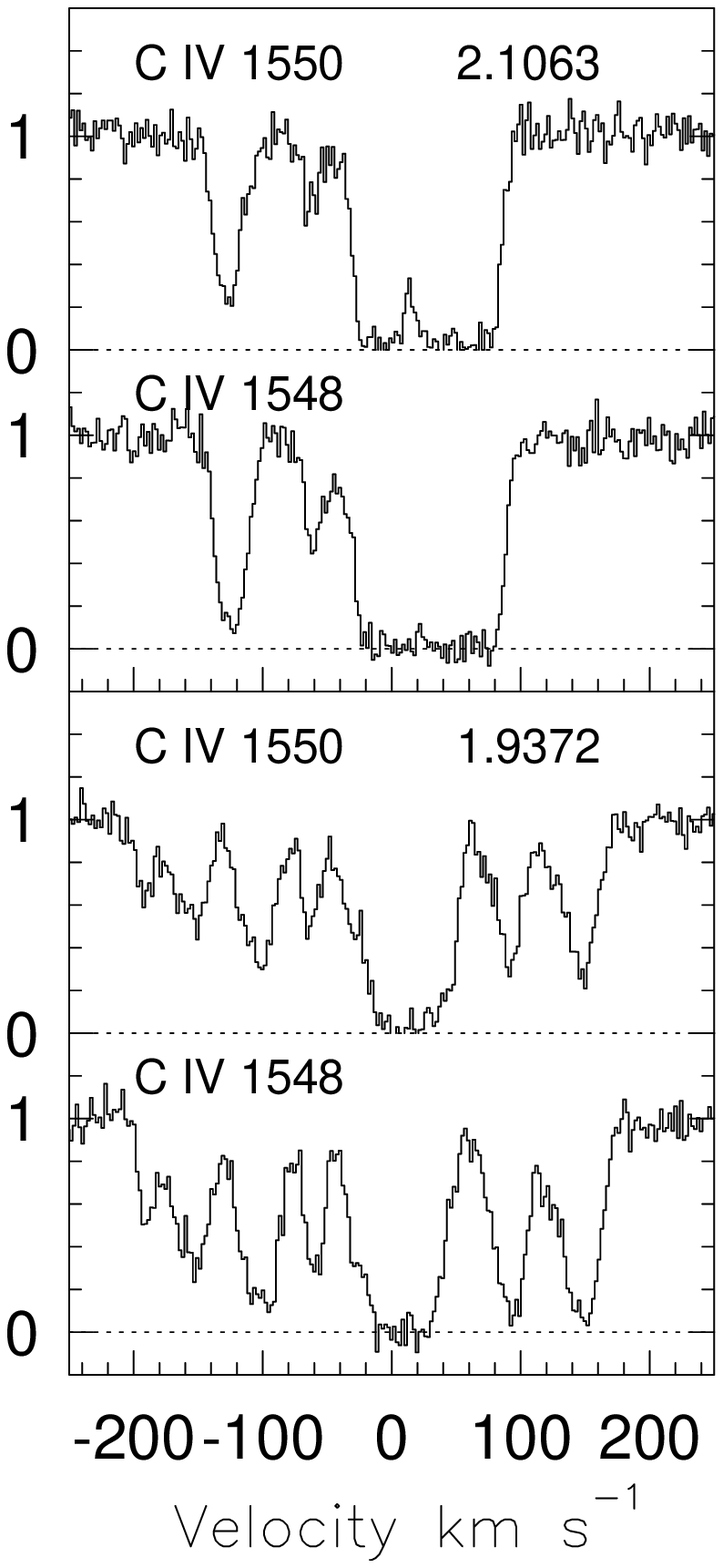}{1.7in}{0}{42.}{42.}{-225}{47}
\vglue -0.85in
\caption{
(right panels) The $z \sim 0.93$ cluster of systems toward QSO
1206+459 (FOS/HST data courtesy D. Schneider).
--- (left panel) An example of resolved {{\rm C}\kern 0.1em{\sc iv}}
profiles.  These HIRES/Keck {{\rm C}\kern 0.1em{\sc iv}} profiles show
each transition of the doublet highly resolved into multiple
subcomponents (upper panels: $z=2.106$; lower panels: $z=1.937$). 
\label{cwcfig:q1206}}
%\vglue -0.22in
\end{figure}

Consider the $z=0.927$ systems shown in the upper right panel of
Figure~\ref{cwcfig:q1206}.
The complex {{\rm Mg}\kern 0.1em{\sc ii}} doublets have a velocity
spread of 500 km~s$^{-1}$.
An additional system is present 1100 km~s$^{-1}$ to the red,
at $z=0.934$.
The lower right panels show the {{\rm C}\kern 0.1em{\sc iv}}, {{\rm
Si}\kern 0.1em{\sc iv}} $\lambda\lambda 1393, 1402$ and {{\rm N}\kern
0.1em{\sc v}} $\lambda\lambda 1238, 1242$ doublets aligned by their
zero point velocities.
The {{\rm C}\kern 0.1em{\sc iv}} doublet has not been resolved, but
the profile suggests complex kinematics.
In the ``CIV1550'' panel, note the previously unreported {{\rm C}\kern
0.1em{\sc iv}} doublet from the $z=0.934$ system.

Photoionization modeling (CLOUDY) of the {{\rm Mg}\kern 0.1em{\sc ii}}
profiles, {{\rm Ly}\kern 0.1em$\alpha$} (not shown), {{\rm C}\kern
0.1em{\sc iv}}, {{\rm Si}\kern 0.1em{\sc iv}}, and {{\rm N}\kern
0.1em{\sc v}} equivalent widths, was unsuccessful at matching the data
(the ionization parameter and  {{\rm H}\kern 0.1em{\sc i}} column
densities were varied for each subcomponent).
Apparently, the systems are not multiple single--phase photoionized
``clouds'' (cf.~\cite{cwcref:bergeron94}).  
The gas could be shock heated (starburst), or the high
ionization gas could be intercluster material not spatially distributed
with the low ionization gas.
An example of how the FOS/HST {{\rm C}\kern 0.1em{\sc iv}} profiles
may appear when resolved is shown in the left panels of
Figure~\ref{cwcfig:q1206}.
Three galaxy candidates, one with a confirmed redshift, have been
identified within 10{\hbox{$^{\prime\prime}$}} of the QSO.
High resolution STIS/HST spectra are required if we are to gain an
appreciation of gas associated with galaxies and their
environments.

A sample of 15 systems have been selected by their HIRES/Keck
{{\rm Mg}\kern 0.1em{\sc ii}} absorption properties for a STIS/HST
study to obtain the first view of the velocity spreads and
cloud--cloud clustering of {{\rm C}\kern 0.1em{\sc iv}} in $z \sim 1$
galactic halos.
This program allows the first direct comparison of {{\rm Mg}\kern
0.1em{\sc ii}} and {{\rm C}\kern 0.1em{\sc iv}} kinematics in $z\sim
1$ galaxies.
Additionally, the $z \leq 1$ {{\rm C}\kern 0.1em{\sc iv}} TPCF from
STIS can be compared to that measured at $z\sim 3.0$
\cite{cwcref:rauch97}, allowing a direct quantification of the
kinematic evolution of high ionization gas.

\section*{The $2.2 \leq z \leq 3.8$ UV Universe}

As seen in the right panel of Figure~\ref{cwcfig:tpcf_dndz}, the
strong {{\rm C}\kern 0.1em{\sc iv}} systems increase in number from
$z=4$ to $z=1.2$.
It has been suggested (see \cite{cwcref:lauroesch96}) that this rapid
evolution is due to an epoch of increased metallicity, and is likely
not due to an evolving UV background flux.
However, the metallicity enrichment scenario is corroborated only by a
few, small, non--uniform data sets of singlet low ionization
transitions, and, as such, is plagued by several uncertainties.  

A uniform sample of {{\rm Mg}\kern 0.1em{\sc ii}} doublets would yield
an unambiguous look at how the low ionization gas evolves out to
$z=4$, providing the leverage needed to settle the ``chemical
enrichment -- UV flux evolution'' debate. 
With an IR spectrograph attached, the Hobby--Eberly Telescope (HET) is
ideally suited for a large {{\rm Mg}\kern 0.1em{\sc ii}} survey.
At Penn State, we (including Beatty, Charlton, Ramsey, \& Schneider)
are building JCAM, an $R = 10,000 - 20,000$ IR spectrograph, for the
HET.  
With JCAM/HET, we can obtain a 0.15~{\AA} 3$\sigma$ rest--frame
equivalent width limit for $2.2 \leq z \leq 3.8$ in $\sim 1$~hr for
a $V=19$ QSO.

With the $R=10,000$ survey, we will obtain the first uniform and
complete sample of low ionization gas to $z=4$.
Our goal is to explore the implications of metallicity enrichment for
galaxy evolution and the redshift evolution of the UV meta--galactic
background.
At $R=20,000$, we will perform follow--up high resolution observations
to study the absorbing gas kinematics.
We aim to construct a $z\sim 3.5$ {{\rm Mg}\kern 0.1em{\sc ii}} TPCF
and directly measure the clustering evolution of low ionization gas by
comparing it to the $z\sim 1$ TPCF measured by Churchill {et.~al}
\cite{cwcref:cvc97}.
We also plan to compare the {{\rm Mg}\kern 0.1em{\sc ii}} TPCF with
the {{\rm C}\kern 0.1em{\sc iv}} TPCF measure by Rauch {et.~al}
\cite{cwcref:rauch97}.
The ultimate goal of our research program is (1) to make the unknown
UV rest--frame known in the currently unexplored redshift regimes,
and (2) to help develop a complete view of the evolution of gas and
its role in galaxy and chemical evolution.

\end{document}